\documentclass[twoside]{dis07}
\usepackage[latin1]{inputenc}
\usepackage[dvips]{graphicx,epsfig,color}
\usepackage{wrapfig,rotating}
\usepackage{amssymb,amsmath,array}
\newcommand{\be}{\begin{equation}}
\newcommand{\ee}{\end{equation}}
\newcommand{\ba}{\begin{eqnarray}}
\newcommand{\ea}{\end{eqnarray}}
\newcommand{\bi}{\begin{itemize}}
\newcommand{\ei}{\end{itemize}}
\newcommand{\ii}{\item}
\def\T{{\cal T}}

\def\g{\gamma}
\def\f{\frac}

\begin{document}
\title{Universality of traveling waves with QCD running coupling}

\author{Guillaume Beuf$^1,$ Robi Peschanski$^1$ and Sebastian Sapeta$^{2,3}$
%
%
\vspace{.3cm}\\
%
%
1- Service de Physique Th\'eorique,
Orme des Merisiers, CEA/Saclay
91191 Gif-sur-Yvette Cedex, FRANCE
\vspace{.1cm}\\
2- Jagiellonian University, Institute of Physics,
Ul. Reymonta 4 PL-30-059 Cracow, POLAND.
\vspace{.1cm}\\
3- Department of Physics, CERN, Theory Division, CH-1211, Geneva 23, 
Switzerland.
\vspace{.1cm}
}

\maketitle

\begin{abstract}

``Geometric scaling'', \emph{i.e.} the dependence of DIS cross-sections on the 
ratio $Q/Q_S,$ where $Q_S(Y)$ is the rapidity-dependent \emph{saturation} 
scale, can be theoretically obtained  from universal ``traveling wave'' 
solutions of the nonlinear Balitsky-Kovchegov (BK) QCD evolution equation at 
fixed coupling. We examine the similar mean-field predictions beyond 
leading-logarithmic order, including running QCD coupling.
\end{abstract}

\section{Motivation}

``Geometric scaling'' (GS) is a striking empirical scaling  property first 
observed in deep-inelastic (DIS) cross-sections 
. It consists in 
the dependence of  $\g^* p$ cross-sections on the 
ratio $Q/Q_S(Y),$ where $\log Q_S \propto Y$ is the rapidity-dependent 
\emph{saturation} 
scale. On a 
theoretical ground, GS can be found as a consequence of saturation effects in 
QCD, when the density of gluons become large enough to impose unitarity 
constraints on the scattering amplitude. It has been shown \cite{munier} that the QCD 
evolution with a nonlinear term describing unitarity damping, the  
Balitsky-Kovchegov (BK) equation, leads to asymptotic ``traveling wave'' 
solutions exhibiting the GS property \cite{munier}. They are ``universal'' 
since they do not depend neither on the initial conditions nor on the precise 
form of the nonlinear damping terms.

\begin{wrapfigure}{r}{0.5\columnwidth}
\centerline{\includegraphics[width=0.45\columnwidth]{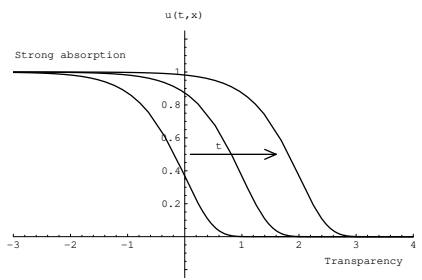}}
\caption{Traveling waves}
\label{1}
\end{wrapfigure}

These results were mainly obtained at leading logarithmic order.
In the present 
contribution, we describe how   higher orders, in particular incorporating 
running QCD coupling, influence these predictions. Potential effects may be due 
to, \emph{e.g.}, next-to-leading (NLL) contributions to the evolution kernel, 
higher-order resummations schemes, observable dependence, infra-red 
regularization, position \emph{vs.} momentum formulation. These aspects and the 
restauration of universality at high enough rapidity $Y$ has been discussed in 
Refs.\cite{us}, whose results are here briefly described. The main difference 
with the fixed coupling prediction is a new kind of geometrical scaling with 
$\log Q_S \propto \sqrt{Y},$ which appears to be as well verified by data 
\cite{geom} as the original GS property.
\section{The Balitsky-Kovchegov equation with  running coupling} 
Before entering the discussion, let us introduce the traveling wave 
method in the case \cite{munier} where the running coupling has been introduced 
\emph{de facto} in momentum space. One writes
\begin{eqnarray}
{ \left[bLog \left(k^2/\Lambda^2\right)\right]}\ {\partial_Y}{\cal T}=
\left\{\chi_{LL}\left(-\partial_{Log k^2}\right)\right\}{\cal T}
-{\cal T}^2\ ,
\label{BK}
\end{eqnarray}
where ${\cal T}(k,Y)$ is the dipole-target amplitude in momentum space, 
$\chi_{LL}$ 
the leading-log QCD kernel and $\left[bLog\left(k^2/\Lambda^2\right)\right]^{-1} 
\!\!\equiv 
\alpha(k^2),$ the one-loop QCD running coupling. The asymptotic solutions of the 
BK equation can be obtained by recognizing the same structure \cite{munier} than 
the  traveling wave equation  $u(t,x)\! \to\! u(t-vx)$
\begin{eqnarray}
{[x]}\ \partial_t u(t,x)=\left\{\partial_x^2 + 1\right\}u(t,x)-u^2(t,x)\ ,
\nonumber
\end{eqnarray}
where the traveling-wave/BK ``dictionnary'' is the following: 

\noindent $Time = t \sim \sqrt Y;$ $Space = x \sim \log k^2;$  \   $ Traveling\ 
wave\ 
u(t,x) =   u(x-v_c t)
\sim {\cal T}.$

\noindent Using the dictionnary, one thus recognizes the GS property  $u(x-v_c 
t) 
\sim {\cal T}(k^2/e^{v_c \sqrt Y}),$ with a saturation scale $Q_S(Y)\sim e^{v_c  
\sqrt Y},$ where $v_c$ 
is the critical wave velocity determined \cite{munier} from the linear kernel 
$\chi_{LL}.$ 

\section{Traveling waves beyond leading QCD logs} 
Let us introduce the general traveling-wave method for the extension beyond QCD 
leading logarithms. It consists in the following steps:
\vspace{-.05cm}
\begin{itemize}
\item Solve the evolution equation restricted to linear terms in terms of a 
dispersion relation: $u(t,x) \sim \int d\g\ e^{\g[x-v(\g)t]}$
\item Find the critical (minimal) velocity $v_c = \min {v(\g)} = v(\g_c)$ which 
is selected by the nonlinear damping {\it independently} of its precise form.
\item Verify sharp enough initial conditions $\g_0 > \g_c,$ in order for the 
critical wave form to be selected.
\end{itemize}
The mathematical properties of such obtained solutions ensure that the 
corresponding asymptotic solutions are ``universal'' that is independent from  
initial conditions, the  nonlinear damping terms and from details of the linear 
kernel away from the critical values. Hence the traveling-wave method defines 
\emph{universality classes} from which different equations admit the same 
asymptotic solutions. 
One caution is that the range of asymptotics may depend on
the singularity structure of the kernel. This may have a phenomenological impact 
on the possibility of using these solutions in the available experimental range 
of rapidities. The saddle-point behaves as $\omega_s \sim Y^{-1/2}$ \emph 
{except} near  singularities in $\g$ of the kernel.

In order to illustrate the method, let us consider the general form of the 
NLL-extended BK equation, replacing in equation \eqref{1}, 
$\chi_{LL}\left(-\partial_{L}\right) \to \chi_{NLL}\left(
-\partial_{L},\partial_Y\right),$ where $L=\log k^2.$ Introducing the function
\be
X(\gamma, \omega) = \int^{\gamma}_{\gamma_0} 
                   d\gamma' \, \chi_{NLL} (\gamma',\omega)\ ,\nonumber 
\ee
 the linear solution reads
\ba 
\T(L,Y)=\int \f{d\g}{2\pi i} \T_0(\g) \ \exp \left[ -\g L + 
\f{1}{b \omega_s} \left( 2 X(\g,\omega_s)-\omega_s \dot{X}(\g,\omega_s) 
\right)\right]\ ,\nonumber \ea
where a dot means  
$\partial_{\omega}$  and  $\omega_s$ is given by the saddle-point equation
\ba \label{eq:cond2}
 Y b \omega_s^2-X(\g,\omega_s)+\omega_s \dot{X}(\g,\omega_s)=0\ .  \ea
From the solutions of the saddle-point  equation, one can infer \cite{us}:
\bi
\ii
 {\it For generic kernels beyond leading logs}: The  
kernels may contain  singularities up to triple poles due
 to the NLL contribution. 
By integration, new single and 
double-pole singularities appear in $X$ at next leading 
order. The universality class is still the same but subasymptotics corrections 
may be large, and thus the critical wave solutions delayed to very large 
energies.
\ii
{\it For Renormalization-Group improved kernels \cite{schemes}}: The  
behaviour of the kernels $\chi_{NLL}$ near the singularities are simple poles. 
This leads only to mild logarithmic singularities in  the 
function $X(\g,\omega).$ The net result \cite{us} is that one finds the same 
universality class as the equation \eqref{1}, since the $\omega$ dependence in 
$X$  can be neglected and thus $\chi_{NLL}\!\to \!\chi_{LL}$. 
\ei
The same approach has been followed for recent QCD formulations of the  
Balitsky-Kovchegov equation with running coupling constant obtained from 
quark-loop calculation \cite{recent}. It leads to the same conclusion (with the 
same warning about eventual kernel singularities): the universality class for the BK equations with 
running coupling is the one defined by Eq.\eqref{1}.

\section{Geometric Scaling in $\sqrt Y.$}
On a phenomenological ground, the main property of  solutions corresponding to 
the universality class of Eq.\eqref{1} is the traveling-wave form $u(x,t) \sim 
u(x-v_c t) = u(k^2/e^{v_c \sqrt Y})$ in the 
\begin{wrapfigure}{r}{0.5\columnwidth}
\centerline{\includegraphics[width=0.45\columnwidth]{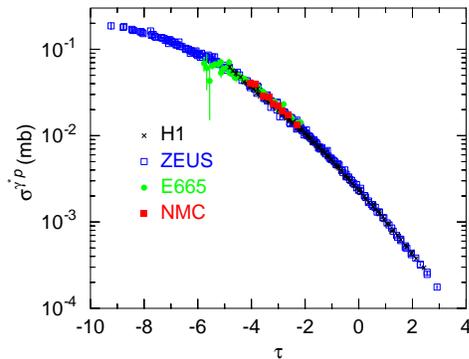}}
\caption{Geometric Scaling in $\sqrt Y$ \cite{geom}}\label{2}
\end{wrapfigure}asymptotic regime.  Assuming a simple relation between that 
amplitude and the  $\g^* 
p$ cross-section, one is led to look for geometric scaling of the form 
$\sigma^{\g^* p} 
\sim (Q^2/e^{v_c \sqrt Y}),$ with $v_c = cst.$ In Fig.\ref{2}, one displays the 
corresponding data plot \cite{geom}. The validity  of the scaling property has 
been quantified using the `Quality Factor'' QF method, which allows to determine 
the adequacy of  a given scaling hypothesis with data independently of 
 the form of the scaling curve \cite{geom}.

\begin{wrapfigure}{r}{0.5\columnwidth}
\centerline{\includegraphics[width=0.45\columnwidth]{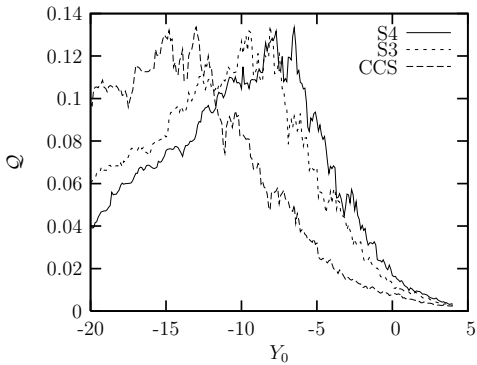}}
\caption{NLL Quality Factors \cite{geom}}\label{3}
\end{wrapfigure}

One may also use the QF method to evaluate the scheme dependence of the 
subasymptotic, nonuniversal terms in the theoretical formulae. In this case, the 
geometric scaling prediction is considered in a ``strong'' version, namely, with 
the critical parameters (such as $v_c$) fixed apriori by the theory. In 
Fig.\ref{3} one displays the QF for geometric scaling for different NLL schemes.
The top QF is larger than $.1$ which ensures a good GS property 
(similar than fig.\ref{2}). Depending on the resummation scheme ($S_3,S_4,CCS,$ 
see \cite{schemes}), $\vert Y_0\vert$ gives  the typical strength of the 
non-universal terms. The S4 scheme seems to reach GS sooner (at smaller 
$\vert Y_0\vert$). 
\noindent  

\section{Conclusions}
Let us give the main results of our analysis:

$\bullet$ \emph {Mean-field saturation beyond leading-logs:}
The modified Balitsky-Kovchegov  equations including  running coupling and  
higher-order QCD corrections to the linear kernel asymptotically converge to the 
same traveling-wave solution. 

$\bullet$  \emph {Characterisation of the universality class:} 
The universality class of these solutions is the BK equation with the leading 
logarithmic BFKL kernel supplemented by  a factorized running coupling whose 
scale is given by the gluon transverse momentum. Higher order contributions to 
the kernel will  affect the subasymptotic behaviour.

$\bullet$  \emph{Higher-order effects in the kernel:}
The renormalization-group improved kernels are expected to improve the 
convergence towards the universal behaviour, spurious singularities being 
canceled.

$\bullet$ \emph {Geometric Scaling:} 
 Geometric scaling in $\sqrt Y$ is a generic prediction of the universality 
class of the BK equation with running coupling. It is well borne out by actual 
data, using the ``Quality Factor'' method \cite{geom} to quantify the validity 
of the scaling hypothesis without assuming  the scaling curve  a priori. 

$\bullet$  \emph{Nonuniversal terms:}
 When using the theoretical ``critical'' parameters geometrical scaling is 
verified but requires the introduction of scheme-dependent subasymptotic

 Prospects of the present studies are interesting. On the theoretical side, it 
would be fruitful to investigate the universality properties of QCD equations 
beyond the mean-field approximation. On the phenomenological side, the problem 
is still not settled to know whether there is a slow drift towards the universal 
solutions or whether it exists subasymptotic traveling wave structures, as 
mathematically \cite{parametric}  or  numerically \cite{albacete} motivated.
  
\begin{footnotesize}



%

\end{footnotesize}
\end{document}